\documentclass[12pt,aps,prd,floatfix,nofootinbib,a4paper,nosuperscriptaddress]{revtex4-2}
\pdfoutput=1

\usepackage{amsmath, amssymb, amsfonts, amsthm, latexsym, epsfig, mathrsfs, xcolor, bbm, slashed, braket, thmtools, cancel}

\usepackage[all]{xy}

\usepackage[inline]{enumitem}

\usepackage{setspace}
\usepackage[marginal, multiple]{footmisc}

\usepackage[T1]{fontenc}
\usepackage[utf8]{inputenc}
\usepackage{lmodern}
\usepackage{array}
\usepackage{makecell}

\usepackage[colorlinks, allcolors=blue!70!black, linktocpage]{hyperref}

\usepackage{standalone}
\usepackage{tikz}
\usetikzlibrary{decorations.pathmorphing,positioning}

\numberwithin{equation}{section}

\usepackage{cleveref}

\usepackage{microtype}

\usepackage{floatrow}

\let\OLDtableofcontents\tableofcontents
\renewcommand\tableofcontents[1]{%
    {\baselineskip 0.5ex %
	\OLDtableofcontents{#1}}%
}

\let\OLDthebibliography\thebibliography
\renewcommand\thebibliography[1]{%
	\setstretch{1.079} 
	\OLDthebibliography{#1}%
	\small %
	\setlength{\itemsep}{0.2\baselineskip} 
}

\let\OLDfootnote\footnote
\renewcommand\footnote[1]{%
	\setlength{\footnotesep}{0.75\baselineskip}%
	{\footnotesize \OLDfootnote{#1}}%
}

\setlist[enumerate]{noitemsep, label=(\arabic*), ref=(\arabic*)}

\renewcommand\thesection{\arabic{section}}
\renewcommand\thesubsection{\arabic{subsection}}

\makeatletter
\def\p@subsection{\thesection.}
\def\p@subsubsection{\thesection.\thesubsection.}
\makeatother 

\theoremstyle{plain}

\theoremstyle{definition}

\declaretheorem[style=remark,qed=$\scriptstyle{\blacksquare}$,numberwithin=section]{remark} 

\creflabelformat{equation}{#2#1#3}

\crefname{section}{sec.}{sec.}
\crefname{appendix}{appendix}{Appendices}
\crefname{figure}{Fig.}{Figs.}
\crefname{table}{Table}{Tables}

\crefname{definition}{Def.}{Defs.}
\crefname{prop}{Prop.}{Props.}
\crefname{lemma}{Lemma}{Lemmas}
\crefname{corollary}{Cor.}{Cors.}
\crefname{thm}{Theorem}{Theorems}
\crefname{remark}{Remark}{Remarks}

\crefname{ass}{Assumptions}{Assumptions}
\crefname{property}{Properties}{Properties}

\newcommand{\be}{\begin{equation}\begin{aligned}}
\newcommand{\ee}{\end{aligned}\end{equation}}

\newcommand{\mc}{\mathcal}

\newcommand{\defn}{\mathrel{\mathop:}=} 

\let\oldint\int
\renewcommand{\int}{\oldint\limits}

\let\oldlim\lim
\renewcommand{\lim}{\oldlim\limits}

\renewcommand{\bar}{\overline}

\newcommand{\Hilb}{\mathscr{H}}

\newcommand{\antiHilb}{%
\hspace{4pt} 
  \vbox{%
    \hrule height 0.5pt
    \kern0.25ex
    \hbox{%
      \kern-0.3em
      \ifmmode\Hilb\else\ensuremath{\Hilb}\fi
      \kern0em
    }
  }
}

\begin{document}

\setstretch{1.2}

\title{Estimating the asymptotics of integer partitions in intermediate dimensions ($d = 3,4,5,6$)}

\author{Avinandan Mondal}
\email{avinandan@alumni.iitm.ac.in}
\affiliation{Raman Research Institute, Sadashivanagar, Bengaluru 560080, India.}

\begin{abstract}
It was recently shown by Yeliussizov \cite{Yeliussizov} that integer partitions in dimensions $d \geq 7$ asymptotically grow strictly faster than MacMahon numbers. As MacMahon numbers match with integer partitions in dimensions $d = 1,2$, the comparison of asymptotics of integer partitions with MacMahon numbers in intermediate dimensions ($d = 3,4,5,6$) is an open question. In this work, we perform Markov chain Monte Carlo (MCMC) simulations till $N=15000$ by using adaptive weight learning followed by conventional MCMC steps to numerically estimate the asymptotics of integer partitions in these intermediate dimensions. We numerically establish that in these intermediate dimensions, partitions asymptotically grow faster than MacMahon numbers. More specifically, assuming that the limits exist, we show: $\lim_{n\to\infty}n^{-3/4}\log p_3(n) = 1.8196 \pm 0.0019$, $\lim_{n\to\infty}n^{-4/5}\log p_4(n) = 1.7215 \pm 0.0045$, $\lim_{n\to\infty}n^{-5/6}\log p_5(n) = 1.6521 \pm 0.0059$, and $\lim_{n\to\infty}\log n^{-6/7}p_6(n) = 1.652 \pm 0.021$ for partitions in dimensions $d=3,4,5,$ and $6$ respectively. These numbers are all larger than MacMahon leading order asymptotic coefficients of $1.7898, 1.6614, 1.5737,$ and $1.509$ respectively. Additionally, we also find estimates for some of the sub-leading asymptotic terms in $\log p_d(n)$ in each of the dimensions. 
\end{abstract}

\maketitle
\newpage 
\tableofcontents

\section{Introduction}\label{sec:intro}

Higher dimensional integer partitions \cite{Andrews, MacMahon, GovNotes} form a class of \textit{easy to state, but hard to solve} interesting problems in combinatorics. It also has deep connections to statistical physics, e.g. the restricted $d$-dimensional partition function of $n\in \mathbb N$ counts the number of microstates at energy $n$ of $q \to \infty$ Potts model in $(d+1)$-dimensions \cite{Potts1, Potts2} \footnote{In the thermodynamic limit, when lattice size is infinite, the microstates are counted by the usual (unrestricted) $d$-dimensional partition function} as well as the number of directed compact lattice animals in hypercubic lattices \cite{Potts1}, plane ($d=2$) partitions and its limit shapes arise in the study of $3D$ Ising model on cubic lattice \cite{Ising}, etc. Integer partitions also appear in different areas in high energy physics, e.g. plane partitions arise in the statistical mechanical model of crystal melting counting BPS states \cite{Ooguri} with the limit shape of the plane partition (which corresponds to the thermodynamic limit for the melting crystal) coinciding with the projection of the shape of mirror Calabi-Yau manifold \cite{Yamazaki}, solid ($d=3$) partitions arise in the \textit{Magnicient Four} model which computes the refined index of a system of $D0$-branes in the presence of $D8-\bar{D8}$ system \cite{Nekrasov}, etc. Despite being so ubiquitous in physics, their behaviour is poorly understood, particularly in higher dimensions. There is no closed form formula for partitions in any dimension \footnote{How there is an exact convergent infinite series for ordinary ($d=1$) partitions due to Rademacher \cite{Rademacher} which was an improvement of the asymptotic formula of Hardy and Ramanujan \cite{HR}. The modularity of $d=1$ generating function is what makes the beautiful machinery of Hardy-Ramanujan-Rademacher work. The $d=2$ generating function being not modular, one cannot use those methods get such exact convergent infinite series as Rademacher \cite{Rademacher} (although non-modularity does not exclude the possible existence of a convergent infinite series expression). However, just the existence of generating function in $d=2$ gives us a lot of analytic power on the asymptotic behaviour of plane partitions, with an exact analytic result for asymptotic behaviour due to Wright \cite{Wright}, (which we use much later in \cref{eq:Wright}).}, while closed form formula for their generating functions exist only in $d=1,2$. Partitions in dimensions $d \geq 3$ are not much well understood theoretically. However, there are some analytic results bounding $d$-dimensional partitions, with some very interesting recent results \cite{Yeliussizov, Oganesyan} which shall be discussed in the main text of the paper. In particular, a very sharp bound by \cite{Yeliussizov} shows that the partitions grow strictly faster than MacMahon numbers asymptotically in all dimensions $d \geq 7$. In this paper, we address the question about what happens in the remaining intermediate dimensions $d = 3,4,5,6$ by performing numerical studies and this will give a complete comparison of leading order asymptotics of integer partitions with MacMahon numbers in all dimensions. In particular, our results coupled with \cite{Yeliussizov} show that partitions grow strictly faster than MacMahon numbers in all dimensions $d \geq 3$.  

\section{Review of some preliminaries}

On $\mathbb N_0^{k}$ ($k \geq 1$), we put a partial ordering by defining $\alpha \equiv (\alpha_1, ..., \alpha_k) < \beta \equiv (\beta_1, ..., \beta_k)$ iff $\alpha_i \leq \beta_i$ $\forall i = 1, ..., k$. Then a $d$-dimensional partition of $n \in \mathbb N$ is finite order ideal $\lambda \subset \mathbb N_0^{d+1}$ such that $|\lambda| = n$, where the cardinality $|\lambda|$ denotes the number of lattice points (i.e. elements of $\mathbb N_0^{d+1}$) in $\lambda$. The set of all $d$-dimensional partitions of $n$ is denoted by $\mc P^{(d)}_n$:
\be
    \mc P^{(d)}_n \defn \{\lambda \subset \mathbb N_0^{d+1} \text{ finite order ideal } \hspace{3mm} | \hspace{3mm} |\lambda| = n\}
\ee
The number of $d$-dimensional partitions of $n$ is the value of $d$-dimensional partition function $p_d$ at $n$:
\be
    p_d(n) \defn |\mc P^{(d)}_n|
\ee
The generating function for $d$-dimensional partitions, $F_d$ is defined by:
\be
    F_d(t) \defn 1 + \sum_{n=1}^{\infty}p_d(t)t^n
\ee
The most analytically well-understood cases are $d=1,2$ where one has a closed form expression for $F_d$ due to Euler and MacMahon respectively: 
\be
    F_1(t) = \big(\prod_{n=1}^{\infty}(1 - t^n)\big)^{-1} \\
    F_2(t) = \big(\prod_{n=1}^{\infty}(1 - t^n)\big)^{-n}
    \label{eq:generating-func}
\ee
MacMahon had conjectured a general expression for generating function which turned out to be wrong for $d \geq 3$ \cite{Atkin}. His expression was of the form:
\be
    M_d(t) = \big(\prod_{n=1}^{\infty}(1 - t^n)\big)^{-\binom{n-d+2}{d-1}} =: \sum_{n=1}^{\infty}m_d(n)t^n
    \label{eq:MacMahon}
\ee
and we only have $F_1 \equiv M_1$ and $F_2 \equiv M_2$. $M_d$ is the generating function of \textit{MacMahon numbers} $m_d(n)$ defined via \cref{eq:MacMahon}. Note that MacMahon numbers and partitions match for all $n \in \mathbb N$ in $d = 1,2$. By virtue of the closed form generating functions in $d =1,2$ (\cref{eq:generating-func}), it is quite easy to exactly enumerate normal ($d= 1$) and plane ($d=2$) partitions. The asymptotic behaviour of MacMahon numbers can be evaluated from the generating function $M_d$ as outlined in \cite{Gov11}: 
\be
    \log m_d(n) \sim \sum_{r=1}^{d}\mu_r^{(d)}n^{\frac{d-r+1}{d+1}} + \nu^{(d)}\log n + \epsilon^{(d)},
    \label{eq:asymp-Mac}
\ee
where $\mu_r^{(d)}, \nu^{(d)}, \epsilon^{(d)}$ are real coefficients and the leading coefficient $\mu_1^{(d)}$ is:
\be
    \mu_1^{(d)} = \lim_{n \to \infty} n^{-\frac{d}{d+1}} \log m_d(n) = \frac{d+1}{d}\big(d \zeta(d+1)\big)^{\frac{1}{d+1}},
\ee
$\zeta$ being the Riemann zeta function. As we do not have a closed form expression for generating function of partitions $F_d$ in $d \geq 3$, analytic understanding of their asymptotics is not yet known completely. It was shown in \cite{Bhatia} that aymptotically $\log p_d(n)$ is bounded between multiples of $n^{d/(d+1)}$:
\be
    \log p_d(n) = \Theta(n^{\frac{d}{d+1}}).
    \label{eq:Bhatia-result}
\ee
In \cite{Yeliussizov}, \cref{eq:Bhatia-result} was made more precise and it was shown that $\log p_d(n)$ is asymptotically bounded for large $n$ as: 
\be
    A^{(d)}n^{\frac{d}{d+1}} \leq \log p_d(n) \leq B^{(d)}n^{\frac{d}{d+1}} + d\log n,
    \label{eq:Damir-bound}
\ee
where $A^{(d)} = \frac{d+1}{(d+1)!^{1/(d+1)}}c^{(d)}$ with some $c^{(d)} > \log 2$ and $B^{(d)} = (d+1)\zeta(d+1)^{1/(d+1)}$.\footnote{Here it is also worth mentioning another quite interesting result due to Oganesyan \cite{Oganesyan} which we shall not use. It states that for extremely large $n$, more specifically for $n \geq (30(d+1))^{2(d+1)^2}$ one has:
\[
1 < n^{-\frac{d}{d+1}}\log p_d(n) < 7200
\]
Note that this bound is very crude compared to \cref{eq:Damir-bound} for low dimensions (e.g. $d=3,4,5,6$) which are of our interest. However, the interesting fact about Oganesyan's result is that the bounds are dimension independent and for very large dimensions, the upper bound performs much better than that of Yeliussizov.} Then subsequently a comparison with $\mu_1^{(d)}$ reveals that:
\be
    A^{(d)} > \mu_1^{(d)} \quad \forall d \geq 7,
    \label{eq:Damir-result}
\ee
and hence in $d \geq 7$, number of partitions $p_d(n)$ is strictly greater than MacMahon numbers $m_d(n)$ for large $n$. However, note that it is not yet rigorously proven that the limit $\lim_{n \to \infty}n^{-\frac{d}{d+1}}\log p_d(n)$ exists for $d \geq 3$. \footnote{Note that \cref{eq:Damir-bound} simply tells us that $n^{-\frac{d}{d+1}}\log p_d(n)$ is bounded between $A^{(d)}$ and $B^{(d)}$ for large $n$ but it does not say that it converges to a value in that range.} Now, inspired by previous works \cite{Bhatia, Gov11, Destainville}, we state the major assumption of this paper as follows:
\begin{itemize}
    \item \textbf{Assumption:} The asymptotic behaviour of $\log p_d(n)$ is given by:
    \be
        \log p_d(n) \sim  \sum_{r=1}^{d}\alpha_r^{(d)}n^{\frac{d-r+1}{d+1}} + \beta^{(d)}\log n + \eta^{(d)} + \cdots,
        \label{eq:assump}
    \ee
    where $``\cdots"$ represents (possible) additional terms that die off at large $n$ and $\alpha_r^{(d)}, \beta^{(d)}, \eta^{(d)}$ are real coefficients. 
\end{itemize}
Note that it was conjectured in \cite{Gov11} that $\alpha_r^{(d)} = \mu_r^{(d)}$, $\nu^{(d)} = \beta^{(d)}$ and $\epsilon^{(d)} = \eta^{(d)}$ for all $d$. \footnote{Recall that it is trivially true for $d =1,2$ as MacMahon numbers and partitions are same in those dimensions. Also, the conjecture was supported by numerical results on asymptotic behaviour of solid ($d = 3$) partitions in \cite{Mustonen} and exact enumeration in \cite{Gov11}.} However an evidence against the conjecture was presented in \cite{Destainville} where it was shown by Monte Carlo simulation that $\alpha_1^{(3)} = 1.822 \pm 0.001$ whereas $\mu_1^{(3)} = 1.7898$. The conjecture in \cite{Gov11} was proven wrong for $d \geq 7$ by virtue of \cref{eq:Damir-result} in \cite{Yeliussizov}. 

The main aim of this paper is to estimate the coefficients in \cref{eq:assump} using Monte Carlo simulations and to compare the leading coefficients $\alpha_1^{(d)}$ obtained from analysis of data generated by the simulation with MacMahon coefficients $\mu_1^{(d)}$ for $d = 3,4,5,6$.

\section{The $d$-dimensional partition graph}

Consider the $d$-dimensional partitions up to some $N \in \mathbb N$ as a graded graph whose vertices are the partitions $\bigcup_{1 \leq n \leq N} \mc P^{(d)}_n$ and the levels are such that the $k$-th level ($1 \leq k \leq N$) contains all $d$-dimensional partitions of $k$. \footnote{Our level indexing starts with $1$ instead of $0$, to match it with the number whose partitions are contained in that level.} The edges of the graph are legal insertions (and hence deletions) of nodes in a partition and the edges are directed towards the partition with larger number of nodes. Clearly level $1$ of the graph is the unique partition of $1$. This is the $d$-dimensional partition graph up to $N$. Generally $N$ is a (large \footnote{Here large means it is quite large compared to the biggest number whose exact $d$-dimensional partition is exactly known}) number whose $p_d(N)$ we are interested at. 

Now, for some $\lambda \in \mc P^{(d)}_n$, define:
\be
    \Gamma_+(\lambda) \defn \{ \mu \in \mc P^{(d)}_{n+1} | \lambda \subset \mu\} \quad (1 \leq n \leq N-1), \\
    \Gamma_-(\lambda) \defn \{ \mu \in \mc P^{(d)}_{n-1} | \lambda \supset \mu\} \quad (2 \leq n \leq N),
\ee
and $\Gamma_+(\lambda) \defn 0$ if $n = N$ and $\Gamma_-(\lambda) \defn 0$ if $n = 1$. We further denote $\gamma_+(\lambda)$ and $\gamma_-(\lambda)$ as the number of outgoing and incoming edges at the partition $\lambda$:
\be
    \gamma_+(\lambda) \defn |\Gamma_+(\lambda)| \\
    \gamma_-(\lambda) \defn |\Gamma_-(\lambda)|
\ee
Now, the total number of incoming edges at level $n+1$ is equal to the number of outgoing edges at level $n$ and hence one has:
\be
    \sum_{\lambda \in \mc P^{(d)}_{n}(\lambda)} \gamma_+(\lambda) = \sum_{\lambda \in \mc P^{(d)}_{n+1}(\lambda)} \gamma_+(\lambda)
    \label{eq:balance}
\ee
Now, if one defines the average number of outgoing and incoming edges for a partition at level $n$ as:
\be
    \langle \gamma_{\pm}(n) \rangle \defn \frac{1}{p_d(n)}\sum_{\lambda \in \mc P^{(d)}_{n}(\lambda)} \gamma_{\pm}(\lambda)
    \label{eq:Imp1}
\ee
then one can write from \cref{eq:balance} that:
\be
    p_d(n+1) = \frac{\langle \gamma_{+}(n) \rangle}{\langle \gamma_-(n+1) \rangle}p_d(n)
\ee
Suppose one knows $p_d(n_0)$, where henceforth we will refer to $n_0$ as the \textit{anchor}. Then one can find $p_d(n)$ for any $n \leq N$ by:
\be
    p_d(n) = \Big(\prod_{k = n_0}^{n-1} \frac{\langle \gamma_{+}(k) \rangle}{\langle \gamma_-(k+1)} \Big)p_d(n_0)
    \label{eq:primary}
\ee
Now, evidently computing all $\langle \gamma_{\pm}(n) \rangle$ exactly for all $n \in [1, N]$ would amount to exactly determining $p_d(N)$. As partitions become extremely large at large n (as evident from \cref{eq:assump}), so we need a way to replace $\langle \gamma_{\pm}(n) \rangle$ exact averages by an average computed over a much smaller set. This kind of replacing an average over an enormous finite set by a random average over a finite set is precisely the purpose of Monte Carlo simulations. 

Before we go further, we rewrite \cref{eq:primary} for $n = N$ as:
\be
    \log p_d(N) = \sum_{k = n_0}^{N-1}\Delta_d(k) + \log p_d(n_0),
    \label{eq:Delta-1}
\ee
where 
\be 
    \Delta_d(k) = \log \langle \gamma_{+}(k) \rangle - \log \langle \gamma_{-}(k+1) \rangle.
    \label{eq:Delta-2}
\ee
In our works, we will use the anchor $n_0$ as the largest value whose exact partition is known as of writing this paper (e.g. it is $72$ for $d=3$) according to OEIS \cite{OEIS}. In the next section, we give a detailed exposition to the algorithm we used for our simulations.

\section{Outline of the algorithm for the Markov chain Monte Carlo simulation}

The algorithm we use for the MCMC simulation to compute the $\Delta_d(k)$ quantities (see \cref{eq:Delta-2}) is not a completely novel method, but rather uses various ideas already present and routinely employed in statistical physics. In particular, the algorithm is a novel fusion of ideas from the Monte Carlo simulations of Mustonen-Rajesh \cite{Mustonen} and Destainville-Govindarajan \cite{Destainville} for solid partitions, while significantly differing from both of them. Very broadly speaking, it is a Monte Carlo method of random walks on a Markov chain whose transition probabilities are fixed by a stationary distribution which is computed by adaptively learning the weights (via Wang-Landau algorithm \cite{WL} followed by Belardinelli-Pereyra $1/t$ steps \cite{1/t}). We will elaborate on this and also compare the method with those in \cite{Mustonen} and \cite{Destainville} in this section. 

\subsection{Theory and motivation: Top-down approach to the algorithm}

Instead of directly providing the algorithm, we will motivate the algorithm in this sub-section. Consider the $d$-dimensional partition graph for some dimension $d$ up to some $N$ which is the largest value whose partition we are interested in. Recall that the goal is to compute $\Delta_d(k)$ which in turn requires the averages $\langle \gamma_{\pm}(k)\rangle$. \footnote{There are methods to exactly traverse the partition graph and compute the partitions directly, e.g. the Bratley-McKay algorithm \cite{BM}, which becomes very computationally expensive. That's the reason why we switch to MCMC simulation instead of exact enumeration.} The Monte Carlo method randomly traverses the partition graph and at each visited vertex (i.e. partition) $\lambda$ it computes $\Gamma_{\pm}(\lambda)$ and then at each level $k$ it estimates $\langle \gamma_{\pm}(k)\rangle$ from the visited vertices at that level. Now the key point is to traverse the graph \textit{randomly}. This means that the probability of the random walker (i.e. the Markov chain) to be found at some partition at some level is uniform across all the levels and uniform across all partitions for any given level. This is not easy to attain as a uniform random selection of a legal growth or deletion at each partition will not produce such a probability distribution for the random walker \footnote{To see this, say one has attained such a probability distribution for the walker. Then, say at level $n$ it is equally likely for the random walker to be at any of the partitions. Now a uniform random selection of a legal growth or deletion will result in an uniform distribution of the random walker at all partitions at levels $n-1$ and $n+1$ if each partition at level $n$ is connected to equal number of partitions at level $n+1$ and $n-1$ via legal growths and deletions respectively. But that is evidently not the case!}. So we instead translate the problem to a Markov chain where the state space is the partition graph (in actual implementation, one splits the graph into overlapping \textit{windows}, where each window $W$ is the graph between levels $n_{W, min}$ and $n_{W, max}$ and we shall consider window width $|W| \defn n_{W, max} - n_{W, min}+1$) and the stationary distribution \footnote{Recall that for a Markov chain, a stationary distribution is the answer to the question: \textit{what is the probability that the random walker is at the state (here, partition) $\lambda$?} Mathematically it is defined as the distribution over the state space such that if $P(\lambda \to \mu)$ is the transition probability from partition $\lambda$ to $\mu$, then $\pi(\mu) = \sum_{\lambda}P(\lambda \to \mu)\pi(\lambda)$ where the sum is over all possible states $\lambda$.} $\pi$ of the Markov chain is such that for any partition $\lambda$:
\begin{enumerate}
    \item \label{item1} the probability that the partition $\lambda$ is a partition of $k \in [n_{W, min}, n_{W, max}]$ is independent of $k$, i.e. $\pi(|\lambda| = k) = \frac{1}{|W|}$, and,
    \item \label{item2} the probability that the partition is $\lambda$ given it is a partition of $k$ is uniform across all partitions of $k$, i.e. $\pi(\lambda | |\lambda| = k) = \frac{1}{p_d(k)}$,
\end{enumerate}
and hence the stationary distribution is:
\be
    \pi(\lambda) = \frac{1}{|W|p_d(k)},
    \label{eq:pi}
\ee
if $\lambda$ is at level $k$. Now, the stationary distribution $\pi$ specified in \cref{eq:pi} is of course not known as we don't know $p_d(k)$ and that's exactly what we are trying to know: so the argument might feel circular as of now! But there is a very good way to actually estimate $\pi(\lambda)$ without directly knowing $p_d(k)$ which we will explain soon. But, let's get ahead of ourselves a bit and say, we have done that and we denote our estimated $\pi$ by the distribution $\hat\pi$. Now given a Markov chain with a stationary distribution and a proposal kernel (i.e. a proposal probability for transition from some state $\lambda$ to $\mu$, $q(\lambda \to \mu)$), one can construct the instantaneous transition probability $P(\lambda \to \mu)$ via Metropolis-Hastings (MH) algorithm. MH algorithm says that a proposal of transition from $\lambda$ to $\mu$ (proposed in accordance to the probability $q(\lambda \to \mu)$) is accepted with a probability $\alpha(\lambda \to \mu)$ given by:
\be
    \alpha(\lambda \to \mu) = \text{min}\Big\{1, \frac{\hat\pi(\mu)q(\mu \to \lambda)}{\hat\pi(\lambda)q(\lambda \to \mu)}\Big\}
    \label{eq:prop-accept}
\ee
and hence the transition probability $P(\lambda \to \mu)$ is: \footnote{For readers who are unfamiliar to the MH algorithm, one can easily show that the transition probability \cref{eq:transition} keeps the distribution $\hat\pi$ stationary. To see that say $\frac{\hat\pi(\mu)q(\mu \to \lambda)}{\hat\pi(\lambda)q(\lambda \to \mu)} < 1$ and hence $\alpha(\lambda \to \mu) = \frac{\hat\pi(\mu)q(\mu \to \lambda)}{\hat\pi(\lambda)q(\lambda \to \mu)}$. Thus, $P(\lambda \to \mu) = q(\lambda \to \mu)\alpha(\lambda \to \mu) = \frac{\hat\pi(\mu)q(\mu \to \lambda)}{\hat\pi(\lambda)}$. Now clearly $\frac{\hat\pi(\lambda)q(\lambda \to \mu)}{\hat\pi(\mu)q(\mu \to \lambda)} > 1$ and hence $\alpha(\mu \to \lambda) = 1$. Thus $P(\mu \to \lambda) = q(\mu \to \lambda)$. Hence one has: $P(\lambda \to \mu) = \frac{\hat\pi(\mu)P(\mu \to \lambda)}{\hat\pi(\lambda)}$. Rearranging one has: $P(\lambda \to \mu)\hat\pi(\lambda) = \hat\pi(\mu)P(\mu \to \lambda)$ which is called \textit{detailed balance}. Now, detailed balance implies stationarity of $\hat\pi$ as: $\sum_{\lambda}P(\lambda \to \mu)\hat\pi(\lambda) = \sum_{\lambda}\hat\pi(\mu)P(\mu \to \lambda) = \hat\pi(\mu)\sum_{\lambda}P(\mu \to\lambda) = \hat\pi(\mu)$.}
\be
    P(\lambda \to \mu) = q(\lambda \to \mu)\alpha(\lambda \to \mu)
    \label{eq:transition}
\ee
Now for the Markov chain over the state space of the window $W$, only those transitions are allowed which involve addition or deletion of a single node from the partition. At every partition $\lambda \in \mc P_k^{(d)}$, one has the following proposal kernel: For $n_{W, min} < k < n_{W, max}$ :
\be
    q(\lambda \to \mu) = \frac{1}{2\gamma_{+}(\lambda)} \quad \text{if } \exists \text{ an edge from } \lambda \text{ to } \mu, \\
    q(\lambda \to \mu) = \frac{1}{2\gamma_{-}(\lambda)} \quad \text{if } \exists \text{ an edge from } \mu \text{ to } \lambda,
    \label{eq:proposal-1}
\ee
whereas at the edges of the window only movement into the window is allowed and hence:
\be
    q(\lambda \to \mu) = \frac{1}{\gamma_{+}(\lambda)} \quad \text{if } k = n_{W, min}, \\
    q(\lambda \to \mu) = \frac{1}{\gamma_{-}(\lambda)} \quad \text{if } k = n_{W, max}
    \label{eq:proposal-2}
\ee
The MCMC algorithm then starts the random walker from some partition at the level $n_{W, min}$ and traverses the window $W$ by dynamically computing the transition probabilities via the MH algorithm \cref{eq:prop-accept} with the proposal kernel described above in \cref{eq:proposal-1} and \cref{eq:proposal-2}. At every partition $\lambda$ it reaches, it counts $\Gamma_{\pm}(\lambda)$ and hence $\gamma_{\pm}(\lambda) = |\Gamma_{\pm}(\lambda)|$. The walker tracks $\gamma_{\pm}(\lambda)$ for each partition $\lambda$ visited at each level $n$, along with the number of times it visits (or \textit{hits}) the level $n$ by $hits(n)$. Note that if a certain proposed transition is rejected then the Markov chain stays at the same partition and same level, but the corresponding $\gamma_{\pm}(\lambda)$ and $hits(n)$ are counted again. At the end of the simulation, it computes estimates of $\langle\gamma_{\pm}(n)\rangle$ as:
\be
    \langle\gamma_{\pm}(n)\rangle = \frac{\sum_{\lambda \text{ visited at level } n} \gamma_{\pm}(\lambda)}{hits(n)}
\ee
and pools the estimates from all the windows and then from independent MCMC runs to finally use it compute $\Delta_d(k)$ using the definition \cref{eq:Delta-2}. 

Now, we come to the important point of how to obtain $\hat\pi$ distribution as an estimate of the true stationary distribution $\pi$ (\cref{eq:pi}). Firstly note that we only need to know the distribution upto an overall multiplicative constant, as evident from \cref{eq:prop-accept} MH step. So we denote the non-normalized $\hat\pi(\lambda)$ as weight $w(\lambda)$ and it must depend only on $|\lambda|$ (as it is equal for all partitions at a given level). Thus:
\be
    \hat{\pi}(\lambda) \propto w(|\lambda|)
\ee
So we have already satisfied \cref{item2} for the distribution $\pi$. Now, ideally for $\pi(\lambda)$ we should have it $\pi(\lambda) \propto 1/p_d(|\lambda|)$ as then $\pi(|\lambda|=k) \propto \sum_{\lambda \in \mc P^{(d)}_k} 1/p_d(k) = 1$ and hence one has satisfied \cref{item1}. But we do not know $p_d(k)$, so instead we write the weight $w(|\lambda|)$ as:
\be
    w(|\lambda|) = \frac{1}{g(|\lambda|)}
\ee
where $g(|\lambda|)$ is known in literature as density of states (DOS) at level $|\lambda|$. The way to estimate $g(|\lambda|)$ is by using Wang-Landau (WL) algorithm \cite{WL} as follows. First, we write everything in log-space (as $g(|\lambda|)$ gets enormous):
\be
    \log w(|\lambda|) = - \log g(|\lambda|)
\ee
and initially set all $g(|\lambda|) = 1$ for $|\lambda| \in W$ in the window. One then starts a random walker at the beginning of the window and uses $w(|\lambda|)$ as $\hat\pi(\lambda)$ in \cref{eq:prop-accept} to take the next step via MH algorithm. Once it visits a level $k$ it updates the DOS as:
\be
    \log g(k) \mapsto \log g(k) + \log f
    \label{eq:WL}
\ee
where $f > 1$ is a modification factor (and initially set to $f_0 = e$). The algorithm also tracks the number of times each level is visited and dynamically computes a histogram of number of times each level visited $H(k)$ vs the level $k$ in the window. At the next step, it uses modified weights to compute MH acceptance probability and this step is repeated. So as the walker traverses the window, it dynamically modifies the weights. After a certain number of steps, the histogram is checked for flatness (i.e. if all levels are approximately equally visited). If at some check it is found that:
\be
    \text{min}_{k \in W}H(k) \geq C \hspace{2pt}\text{max}_{k \in W} H(k)
    \label{eq:hist-flat}
\ee
where $0 < C < 1$ is a flatness criterion (we took $C = 0.95$ for our simulations) then we reset the histogram, change the modification factor $f$ to $\sqrt{f}$ and re-run. After each run the estimates $g(k)$ get better and better. This is essentially the WL algorithm. Note that the WL algorithm after each iteration tends $f \to 1$ and hence the modification factor becomes smaller and smaller for $\log g(k)$ at each $k$. However it was argued in \cite{1/t} that tis conventional WL algorithm with $\log f \mapsto \frac{1}{2}\log f$ can lead to saturation of error for DOS estimates. Hence, after a certain value of $\log f$ is reached (along with some other conditions are met), we transition to Belardinelli-Pereyra style $1/t$ steps \cite{1/t} where at the first step it sets $\log f = \frac{|W|}{t_0}$ where $t_0$ is the time-step at which the $1/t$ steps start (note that it includes conventional WL steps). Then, at each time step, one sets the modification factor $\log f = \frac{|W|}{t}$ where $t$ is the total number of time steps since the beginning of WL traversal (again including the conventional WL steps) and uses it to modify the DOS estimates as in \cref{eq:WL}. This $1/t$ process is repeated many times until the final value of $\log f$ is small enough (in our runs, we repeated until $\log f \sim 10^{-7}$). Note that \cite{1/t} states that in general better estimates for DOS at the same value of $\log f$ is attained using $1/t$ algorithm after certain number of conventional WL steps instead of using conventional WL entirely. After the $1/t$ steps are completed, we freeze the weights and use these learned weights $w(|\lambda|)$ as the estimate of non-normalized stationary distribution $\hat\pi$ and start the MCMC traversal using MH algorithm as we discussed earlier in this subsection.

\begin{remark}
    \textbf{How our algorithm differs from \cite{Mustonen} and \cite{Destainville}?} Solid partitions have been enumerated before using Monte Carlo simulations by Mustonen \& Rajesh (MR) \cite{Mustonen} as well as by Destainville \& Govindarajan (DG) \cite{Destainville}. In \cite{Mustonen}, the authors followed WL algorithm for DOS estimation and reported the obtained weights (after appropriate normalization at some known value of partition which fixes the multiplicative constant) as $p_3(n)$. So, this is essentially the first part of stationary distribution estimation part of our algorithm (i.e. the conventional WL steps before $1/t$ steps). Reporting the obtained weights after normalization as partition numbers has certain disadvantages. This is because Wang-Landau algorithm does not guarantee convergence of estimates of DOS to true DOS value and there is an error saturation (even after $1/t$ steps) \cite{Bela16}. Hence in our algorithm, WL (along with $1/t$ steps) act like a sampling device to give an approximate stationary distribution which lets the subsequent fixed weight Markov chain Monte Carlo random walkers to traverse the partition graph (or window) in an approximately uniform fashion and compute $\langle \gamma_{\pm}(k)\rangle$ and use it in exact combinatorial formulae \cref{eq:Delta-1} and \cref{eq:Delta-2} to compute $\Delta_d(k)$ and hence $p_d(k)$. Now, coming to the approach in \cite{Destainville}, the authors used the exact combinatorial identities \cref{eq:Delta-1} and \cref{eq:Delta-2} and traversed the partition graph using transition matrix Monte Carlo simulation with fixed weights. However their fixed weights was not obtained by learning adaptively using conventional WL and $1/t$ steps and then freezing them (as in our algorithm). Instead, they used their weight to be $w_{DG}$:
    \be
        w_{DG}(k) = \exp(-\beta k)
    \ee
    for some abstract temperature $\beta$. However, that will not give a stationary distribution satisfying \cref{item1} as in that case the corresponding $\pi(|\lambda|=k) \propto \sum_{\lambda \in \mc P^{(d)}_k} \exp(-\beta k) = p_d(k)\exp(-\beta k) \propto \exp(\alpha_1^{(d)}k^{d/d+1})\exp(-\beta k)$ where the last proportionality is at leading order in $k$, and even at that leading order, it is $k$ dependent (even if someone is able to tune $\beta \approx \alpha_1^{(d)}$). Also note that the Monte Carlo simulation of DG was for restricted partitions in a box and they had estimated the box size to be large enough so that the error due to restriction is negligible. However, we compute unrestricted partitions directly.  
\end{remark}

\subsection{The algorithm: Bottom-up}

Since we have reasoned through what must be the structure of our algorithm, we now collect all the pieces and present it in a time-ordered fashion:
\begin{enumerate}
    \item Divide the partition graph between levels $[n_0-K, N]$ into several overlapping windows. Here $K = 0$ if one does not want to pad the first window. However for long runs, one can pad the lower window to have better statistics. Now consider one such window $W$ of width $|W|$. 
    \item \label{WL} For each level $k \in W$, let $g(k)$ be the density of states (DOS) with $w(k) = \frac{1}{g(k)}$ being the corresponding weight. First set $\log g(k) = 0$ $\forall k \in W$. Then begin a random walk at the beginning of the window $W$. For any partition $\lambda$, use \cref{eq:proposal-1} or \cref{eq:proposal-2} as proposal kernel and for acceptance probability $\alpha(\lambda \to \mu)$ use:
    \be
        \alpha(\lambda \to \mu) = \text{min}\Big\{1, \frac{w(|\mu|)q(\mu \to \lambda)}{w(|\lambda|)q(\lambda \to \mu)}\Big\}
        \label{eq:acc}
    \ee
    Keep track of the number of steps ($t$) and number of times a level $k$ is visited by $H(k)$. For each time a level $k$ visited, change the DOS as:
    \be
        \log g(k) \mapsto \log g(k) + F
        \label{eq:DOS-estimates}
    \ee
    where initially $F$ is set as $F_0 = 1$ set initially ($F = \log f$ of \cref{eq:WL}). After certain number of steps $Z$, check appropriate flatness of the histogram (for our simulations, we took $Z = 10000$). If \cref{eq:hist-flat} is satisfied and each level in the window is visited at least $D$ times (we took $D = 3000$ for our simulations), reset only the histogram to zero values, change the modification factor to $F$ by half $F_{i+1} = \frac{F_i}{2}$ and repeat this step until the $1/t$ step is triggered by \cref{eq:BP-trigger}. 

    \item \label{eq:BP-trigger} After every halving stage $F_{i+1} = F_i/2$, check:
    \begin{itemize}
        \item if $F_{i+1} < 1/2^{x}$ where $x \in \mathbb N$ is minimum number of conventional WL steps (generally taken as $x=16$ in our runs);

        \item if $t \geq |W|$;

        \item if $F_{i+1} \leq |W|/t$.
    \end{itemize}
    If all of them are satisfied, we exit the conventional WL steps and enter $1/t$ steps as follows. We set $F = |W|/t$ after each time-step and use it to update the DOS estimates as in \cref{eq:DOS-estimates}. This step is repeated $M$ times, where $M$ is large enough to make $F$ appropriately small (for our simulations we take $M = 60,000,000$ which leads to final $F \sim 10^{-7}$). Once $M$ steps are completed, the weights $g(k)$ for $k \in W$ are frozen. 

    \item \label{prod} The frozen weights are used now to traverse the window $W$ (we call this as \textit{production phase}). First we traverse the window for some time (using the MH algorithm with fixed weights) to prevent a bias of a beginning point for $y$ steps (this $y$ is called \textit{burn-in steps} and for our simulations we took $y = 10,000,000$). For each level $k \in W$, keep track of three quantities $hits(k)$, $E_+(k)$ and $E_-(k)$ for the number of times level $k$ is visited, sum of $\gamma_{+}(\lambda)$ for all $\lambda$ visited at level $k$ and sum of $\gamma_{-}(\lambda)$ for all $\lambda$ visited at level $k$ respectively. Start from the beginning of the window, use the proposal kernel \cref{eq:proposal-1} or \cref{eq:proposal-2} with acceptance probability as \cref{eq:acc} with weights as fixed by the end of step \cref{eq:BP-trigger}. For each partition $\lambda$ visited at each level $k \in W$, increase: 
    \be 
        hits(k) = hits(k) + 1, \\ E_+(k) = E_+(k) + \gamma_{+}(\lambda), \\ E_-(k) = E_-(k) + \gamma_{-}(\lambda).
    \ee
    Note that these increments must happen even if the transition proposal is rejected at some step and the Markov chain stays at the same partition at that step. Continue this Markov chain random walk $M'$ number of times (for our simulations we took $M' = 60,000,000$ and we call $M'$ number of steps as \textit{production steps}). 

    \item \label{final} Do the steps \cref{WL}-\cref{prod} for all the windows in $[n_0 - K, N]$. Then repeat them over for all the independent runs. At the end of all the independent runs, compute:
    \be
        \langle \gamma_{\pm}(k)\rangle = \frac{E_{\pm}(k)}{hits(k)}
    \ee
    for all $k \in [n_0, N]$ and hence compute $\Delta_d(k)$ by \cref{eq:Delta-2}.
     
\end{enumerate}

Calculations in steps \cref{WL}-\cref{prod} in separate windows can be done parallely and this can drastically reduce wall time for computation when the program is parallelized over a multi-core CPU.

\section{Simulations and Data}

We implemented the algorithm in C and used OpenMP for parallelization. We did the simulations for partitions in dimensions $d = 2,3,4,5,6$ where the plane partition ($d=2$) case is for \textit{algorithm validation} of our algorithm, i.e. to show that our simulations (and subsequent data analysis) give accurate results. The source codes are freely available under the GNU Affero General Public License v3 (GNU AGPLv3) in \cite{Me}. May the source be with you!

For our simulations, in each dimension we simulated till $N=15000$, while we took histogram flatness parameter as $C = 0.95$, minimum number of visits per level in each conventional WL halving of modification factor as $D = 3000$ \footnote{The only exception is for $d=2$ it was taken $D = 1000$. This was not intentional and it was realized after the scientific runs were done. However as $d=2$ was a algorithm validation case anyway, we did not do a separate run with a higher value of $D = 3000$ whose job is to make histogram even flatter in conventional WL stage (leading to a generally better DOS estimate)}, minimum number of conventional WL steps as $x = 16$, number of Belardinelli-Pereyra $1/t$ steps as $M = $ 60 million, number of burn-in steps as $y = $ 10 million, and number of production steps as $M'=$ 60 million for each independent run. The number of independent runs for each dimension was $15$. The generated data is freely available under Open Data Commons Open Database License v1 (ODC ODbL v1.0) in \cite{Me2}. The total wall time taken for the entire batch of $15$ independent runs for dimensions $d=2,3,4,5,6$ are about $2.8$, $3.5$, $5$, $5.8$, and $6.6$ hours respectively when parallelized with 24 OpenMP threads in an AMD Ryzen 9 5000 series CPU.

\section{Analysis of Data}

We have from \cref{eq:assump} that asymptotically (for large n):
\be
    \log p_d(n) =  \sum_{r=1}^{d}\alpha_r^{(d)}n^{\frac{d-r+1}{d+1}} + \beta^{(d)}\log n + \eta^{(d)}    
\ee
Now the MCMC simulation naturally gives $\Delta_d(n)$ as observables and adding them up to obtain $\log p_d(n)$ from the anchor $n_0$ will make the errors grow as $n$ increases. Hence we fit $\Delta_d(n)$ directly into the asymptotic model:
\be
    \Delta_d(n) =  \sum_{r=1}^{d}\alpha_r^{(d)}\big(n^{\frac{d-r+1}{d+1}} - (n-1)^{\frac{d-r+1}{d+1}}\big) + \beta^{(d)}\big(\log (n) - \log (n-1)\big),
    \label{eq:asymptotic}
\ee
and extract the coefficients $\alpha_r^{(d)}$ and $\beta^{(d)}$ by linear regression in the range $[n_{min}, N]$ where $N$ is the maximum value in the data (for us $N = 15,000$) and $n_{min}$ is appropriately chosen. Before that, we do a proper analysis of the sources of uncertainty to accurately put error bounds on our reported values of these coefficients. 

\subsection{Sources of uncertainty}

There are three different sources of uncertainty: 
\begin{enumerate}
    \item \label{sderror} \textbf{Error in $\Delta_d(n)$ due to finite number of MCMC runs:} Different runs give slightly different values of $\Delta_d(n)$. Note that our final $\Delta_d(n)$ at the end of all runs is not the arithmetic mean of individual run $\Delta_d(n)$ values but rather it is computed from pooled values of $E_{\pm}(n)$ as described in algorithm step \cref{final}. However the individual variations of $\Delta_d(n)$ in each run gives an estimate of MCMC error and hence we take its statistical error and call it $\delta_{\Delta}(n)$. Note that $\delta_{\Delta} \sim 1/\sqrt{(\text{number of runs})}$ and hence it gets arbitrarily small for large number of independent runs. 

    \item \label{err:asymp} \textbf{Contamination from pre-asymptotic finite terms in \cref{eq:asymptotic}:} We want to fit an asymptotic model \cref{eq:asymptotic} to our dataset. This inherently has error coming from the pre-asymptotic finite terms (which go to zero at large $n$). 

    \item \label{err:cond} \textbf{Loss of conditioning sensitivity due to finite fit range:} This kind of error happens as linearly independent functions can become \textit{approximately linearly dependent} in a finite interval, leading to trade-offs between different parameter values. To illustrate this, consider the basis functions of the model \cref{eq:asymptotic} for $d=2$ and take their leading $n$ behaviour:
    \be
        \phi_1(n) = \big(n^{\frac{2}{3}} - (n-1)^{\frac{2}{3}}\big) = \big(\frac{2}{3}\big)n^{-\frac{1}{3}} + ... \\
        \phi_2(n) = \big(n^{\frac{1}{3}} - (n-1)^{\frac{1}{3}}\big) = \big(\frac{1}{3}\big)n^{-\frac{2}{3}} + ... \\
        \phi_3(n) = \big(\log (n) - \log (n-1)\big) = n^{-1} + ...
    \ee
    Now write $n = N(1 - z)$ where $z \in [0, 1- \frac{n_{min}}{N}]$ and hence:
    \be
        \phi_1(n) = \big(\frac{2N^{-\frac{1}{3}}}{3}\big)(1 - z)^{\frac{-1}{3}} + ... = \big(\frac{2N^{-\frac{1}{3}}}{3}\big)\big(1 + \frac{1}{3}z + \frac{2}{9}z^2 + ... \big) + ... \\
        \phi_2(n) = \big(\frac{N^{-\frac{2}{3}}}{3}\big)(1 - z)^{\frac{-2}{3}} + ... = \big(\frac{N^{-\frac{2}{3}}}{3}\big)\big(1 + \frac{2}{3}z + \frac{5}{9}z^2 + ... \big) + ... \\
        \phi_3(n) = N^{-1}(1-z)^{-1} + ... = N^{-1}(1 + z + z^2 + ...) + ... 
        \label{eq:z}
    \ee
    Now for small values of $z$ (and the effect will be more if one takes $n_{min}$ to be large), the $z^2, z^3, ...$ terms will be small and only $1,z$ will act as basis functions. Being only two linearly independent dominant functions ($1$ and $z$), the three basis functions $\phi_1, \phi_2, \phi_3$ will become (approximately) linearly dependent and hence finding their coefficients from regression becomes an (approximately) ill-conditioned problem. This problem will amplify in higher dimensions as they have more basis functions, e.g. $d=6$ has $7$ basis functions in \cref{eq:asymptotic}. In fact in higher dimensions, this problem gets so severe that we only restrict our attention in fitting only leading three or two basis functions. 
    
\end{enumerate}

\begin{remark}
    \textbf{Assumption in error \cref{sderror}:} We assumed for our error analysis that $\Delta_d(n)$ produced by individual independent runs are un-correlated. This is generally not true, but we will take the standard deviation as the leading source of error from the MCMC runs. 
\end{remark}

\begin{remark} \label{eq:tricky-n-min}
    \textbf{Uncertainties from \cref{err:asymp} and \cref{err:cond} compete between each other:} To reduce the error due to finite $n$ pre-asymptotic contributions in the asymptotic model \cref{eq:asymptotic} as described in \cref{err:asymp}, one should take $n_{min}$ to be larger (and not just set $n_{min} = n_0$). But taking $n_{min}$ large makes the fit window smaller (in particular as $z \in [0, 1- \frac{n_{min}}{N}]$, so the maximum value of $z$ comes closer to $0$) and amplifies the error \cref{err:cond}. 
\end{remark}

\subsection{Our data analysis strategy}
\label{sec:data-strategy}

As evident from \cref{eq:tricky-n-min}, finding a single $n_{min}$ and approximating the errors due to \cref{err:asymp} and \cref{err:cond} can be very tricky. So we adopt a strategy as follows based on scaled condition number ($\kappa$) for model fit using linear regression: \footnote{We briefly explain scaled condition number in this footnote. The vector $\left( \alpha_1^{(d)},\ldots,\alpha_d^{(d)},\beta^{(d)}
\right)^T$ is the vector of fitted coefficients and $J$ is the design matrix whose columns are the corresponding basis functions evaluated at the data points. Since the data has non-uniform MCMC errors \(\sigma_\Delta(n)\), we first form the weighted design matrix
\[
A=W^{1/2}J,
\qquad
W_{nn}=\frac{1}{\sigma_\Delta(n)^2}.
\]
The weighting takes into account the fact that data points with smaller error carry greater information in the regression. We then normalize the columns of the weighted design matrix $A$ to unit Euclidean norm to get normalized design matrix $\widetilde{A}$. Then define the normalized condition number as:
\[
\kappa \defn 
\frac{s_{\max}(\widetilde A)}
{s_{\min}(\widetilde A)}
\]
where \(s_{\max}(\widetilde A)\) and \(s_{\min}(\widetilde A)\) are the largest and smallest singular values of $\widetilde A$ respectively. A large scaled condition number therefore indicates that there exists a direction in parameter space in which small perturbations of the data can produce comparatively large changes in the fitted coefficients. In the present problem this occurs because, when the fitting interval becomes sufficiently narrow, the asymptotic basis functions become approximately linearly dependent (as explained in \cref{err:cond}). Note that $\kappa$ increases as the number of fitting basis functions increase or as the fitting range decreases.}
\begin{enumerate}
    \item \label{scan} Start scanning from $n_{min} = n_0+1$ and proceed till $n_{min} = n_1$ when $\kappa = \kappa_0$ is reached. The value of upper cut-off to the scaled condition number $\kappa_0$ is set so that it is the least number which makes $n_1 \sim 900-1000$, but at the same time we impose a hard limit of $\kappa_0 \leq 200$. The reason behind keeping $n_1 \sim 900-1000$ is that it is quite deep into large $n$ for a $N=15000$ dataset, thus minimizing the error due to pre-asymptotic terms, while at the same time $n_1>1000$ will discard too much information (exception to this $n_1 \sim 900-1000$ rule is allowed in some circumstances, see the discussion after \cref{rem:rem}). Also keeping the scaled condition number low is important to keep the problem well-conditioned. We keep $200$ as the hard upper cut-off to the scaled condition number as in our dataset, $\delta_{\Delta}(n)/{\Delta(n)} \sim 10^{-4}$ (across $d=2,3,4,5,6$) and hence $\kappa_0 = 200$ fixes the scale of worst case amplification of Monte Carlo noise to about $2\%$ in the fitted parameter vector. Now, store the values of the coefficients and their fitting errors. 
    \begin{remark} \label{rem:rem}
        \textbf{All of the coefficients in \cref{eq:asymptotic} can't be estimated by this strategy:} More basis functions lead to more ill-conditioning of the fit, as we already mentioned in \cref{err:cond}. In fact $\kappa_0 \leq 200$ criterion in \cref{scan} in our strategy cannot be satisfied even for $n_{min} = n_0+1$ for too many basis functions. Thus, only leading few terms shall be fit for higher dimensions. 
    \end{remark}
    However, if while fitting data, it is observed that fitting $m$ basis functions keeps $\kappa \leq 200$ for a few hundred values of $n_1$ (but not till $900-1000$), while fitting $m-1$ basis functions can push $n_1$ to $900-1000$ range for a very small value of $\kappa$, then we need to be careful. In that case, we will perform a goodness of fit at $n_{min} = n_0+1$ for both $m$-parameter and $m-1$ parameter model and choose the one with lower reduced chi-squared value. If goodness of fit favours the $m$-parameter model then we will choose that and disregard our $n_1 \sim 900-1000$ rule. 
    
    \item Take the mean value of $\alpha$ in the range $n_{min} \in [n_0, n_1]$. Then take $\delta_1\alpha = \frac{1}{2}(\text{max. value of } \\ \alpha \text{ in } n_{min}\in[n_0,n_1] - \text{min. value of }\alpha \text{ in } n_{min}\in[n_0,n_1])$ as the error due to finite fit-range. We then take $\delta_2\alpha = \text{max. value of fit error for } n_{min} \in [n_0,n_1]$ as the statistical error from MCMC runs and subsequent fit. We then report $\delta_1\alpha+\delta_2\alpha$ as a conservative estimate of the total error. 
\end{enumerate}

\section{Results}

We now apply our data analysis strategy \cref{sec:data-strategy} to our datasets. Also, we keep $2$ significant digits for all the errors and approximate our reported central values to same number of decimal places. We shall also refer to the exact dataset in \cite{Me2} used in the analysis. The naming convention of the dataset is as follows: the datasets are CSV files with name MC-kD-a-b-c-delta.csv where k is the dimension of the partition, a is the number of independent MC runs to generate the data, b is the anchor $n_0$ and c is $N$, the maximum level in the partition graph used for the simulations. 

\subsection{$d=2$ case (algorithm validation)}

We took $n_0 = 40$ for $d=2$. \footnote{Note that one can easily compute plane partitions exactly for arbitrarily large numbers as one has an exact generating function (as $M_2 \equiv F_2$), so we took $n_0 = 40$ to keep it of roughly the same value as $n_0$ for other dimensions where there is no other way to know partitions than expensive explicit enumerations and hence partitions till some $n_0$ is only known.} We take $\kappa_0 = 70$ which keeps $n_1 = 979$. The results of analysis from the data MC-2D-15-40-15000-delta.csv are: 
\be
    \alpha_1^{(2)} = 2.00946 \pm 0.00049\\
    \alpha_2^{(2)} = 0.003 \pm 0.035\\
    \beta^{(2)} = -0.708 \pm 0.094
    \label{eq:2d-result}
\ee
Also the reconstructed value of plane partition for $N=15000$ is obtained as:
\be
    p_2(15000) = (1.82 \pm 0.83) \times 10^{527}
\ee
Now, Wright's asymptotic result for plane partitions \cite{Wright} from MacMahon generating function (which matches with plane partition generating function) states that theoretically (approximated to same number of decimal places as the fitted coefficients \cref{eq:2d-result}):
\be
    \alpha_{1, \text{theory}}^{(2)} \defn \mu_{1}^{(2)} =  \frac{3}{2}(2\zeta(3))^{1/3} \approx 2.00945\\
    \alpha_{2, \text{theory}}^{(2)} \defn \mu_{2}^{(2)} =  0 \\
    \beta_{\text{theory}}^{(2)} \defn \nu^{(2)}  = -\frac{25}{36} \approx -0.694
    \label{eq:Wright}
\ee
and also from the generating function one can compute $p_2(15000)$ exactly and up to $2$ significant digits after decimal the value is:
\be
    p_{2, \text{theory}} = 1.79 \times 10^{527}
\ee
So, the deviation of our estimated values of $\alpha_1^{(2)}, \alpha_2^{(2)},$ and $\beta^{(2)}$ from the theoretical value turns out to be $Z = 0.02\sigma, 0.09\sigma$, and $0.15\sigma$ respectively. Also, the deviation of our estimated value of plane partition at $N = 15000$ from the theoretical value is $Z = 0.04 \sigma$. Thus we have an excellent agreement with the theory.

\subsection{$d=3$ case}

There are $4$ basis functions in \cref{eq:asymptotic} in $d=3$. However if we try to fit all the $4$ basis functions to the data MC-3D-15-72-15000-delta.csv, the value of $\kappa$ even at $n_{min}=n_0+1=73$ is $214.8$. Hence, we fit the leading $3$ basis functions. We take $\kappa_0 = 120$ which keeps $n_1=918$. The results are:
\be
    \alpha_1^{(3)} = 1.8196 \pm 0.0019 \\
    \alpha_2^{(3)} = 0.155 \pm 0.046 \\
    \alpha_3^{(3)} = -0.39 \pm 0.35
    \label{eq:3d-result}
\ee
Now from MacMahon generating function $M_3$ one has up to 4 significant digits after decimal:
\be
    \mu_1^{(3)} = 1.7898
\ee
We have:
\be
    \alpha_1^{(3)} > \mu_1^{(3)}
\ee
and the deviation is $Z = 15.7\sigma$.

\begin{remark}
    \textbf{Comparison with \cite{Mustonen} result:} MR had obtained an estimate of $\alpha_{1, MR}^{(3)} = 1.79 \pm 0.01$ \cite{Mustonen} which deviates from our result at $Z = 2.96\sigma$. Note that their suggestion that asymptotics of solid partitions should match with that of MacMahon numbers in $d=3$ is wrong. 
\end{remark}

\begin{remark} \label{rem:part-1}
    \textbf{Comparison with \cite{Destainville} result:} DG had obtained all the $4$ coefficients (and had also estimated $\eta^{(3)}$ (see \cref{eq:assump})). Their estimates of the four coefficients are:
    \be
        \alpha_{1, DG}^{(3)} = 1.822 \pm 0.001 \\
        \alpha_{2, DG}^{(3)} = 0.06 \pm 0.03 \\
        \alpha_{3, DG}^{(3)} = 1.0 \pm 0.4 \\
        \beta_{DG}^{(3)} = -0.8 \pm 0.3
    \ee
    Thus, their results of three leading coefficients deviate from ours at $Z=1.12\sigma, 1.73\sigma$, and $2.62 \sigma$. Note that the estimates of DG were obtained for a fit between $n \in [50, 10100]$. Hence they estimated the $4$ coefficients by the least conditioning number possible for their data and then took account of the pre-asymptotic terms error by fitting a pre-asymptotic term and using that to obtain the error-bars, although they reported the values for only the asymptotic expression fit (\cref{eq:asymptotic}). Hence their strategy is quite different from ours. However, for the sake of another comparison we fit our data MC-3D-15-72-15000-delta.csv in the maximum possible range $n\in [73,15000]$ available to us (i.e. setting $n_{min} = n_0+1 = 73$) to the full asymptotic formula (and completely disregarding our data analysis strategy) to obtain the following fit and we report only statistical errors (so quite under-estimated errors):
    \be
        \alpha_{1, n_0 = 72}^{(3)} = 1.822643 \pm 0.000084 \\
        \alpha_{2, n_0 = 72}^{(3)} = 0.0511 \pm 0.0024 \\
        \alpha_{3, n_0 = 72}^{(3)} = 1.114 \pm 0.028 \\
        \beta_{n_0 = 72}^{(3)} = -0.876 \pm 0.013
        \label{eq:3d-2}
    \ee
    and these results agree quite well with DG results. However, since fitting with all $4$ basis functions lead to a relatively ill-conditioned problem and also fitting right from $n_{min} = n_0 = 72$ amplifies the error due to pre-asymptotic terms contaminating the asymptotic formula, so we do not report these numbers as our estimates and report \cref{eq:3d-result} as our result. However, we agree with the broad conclusion of DG that at $d=3$, partitions grow strictly faster than MacMahon numbers. 
\end{remark}

\begin{remark}
    \textbf{Comparison with \cite{Destainville} result: Part-II: Oscillations?} DG had reported a very curious behaviour for the residual:
    \be
        \delta(n) \defn [\Delta_3(n)]_{obs} - [\Delta_3(n)]_{fit} 
    \ee
    where $[\Delta_3(n)]_{obs}$ are the raw values obtained from MCMC simulation and $[\Delta_3(n)]_{fit}$ are the fitted values. They detected an oscillation for $n\delta(n)$ with $n^{1/4}$ as the oscillatory variable. Now, as our result \cref{eq:3d-result} does not give $\beta^{(3)}$ so our analysis pipeline cannot check this result directly. However, in \cref{rem:part-1} we did a fit for $n \in [73,15000]$ and obtained values for all of the coefficients in \cref{eq:3d-2} which also matched quite well with DG results. Hence we used this \cref{eq:3d-2} to detect such an oscillation in the residuals from our data. We first fit $f(n) = A\cos(k n^{1/x} + \phi)$ to $\delta(n)$ vs $n$ from our data over $0.1 \leq x \leq 10$ at an interval of $0.1$ to find the value of $x$ which gives least value for the reduced chi-squared of the fit. This gave several narrow almost equally deep local minima. However, when we fit $f(n) = A\cos(k n^{1/x} + \phi)$ to $n\delta(n)$ vs $n$ from our data over $0.1 \leq x \leq 10$ at an interval of $0.1$ and again find the value of $x$ which gives least value for the reduced chi-squared of the fit, we find a much cleaner result. We now find a broad global minimum at $x=4.3$ and hence the fit \cref{eq:3d-2} does predict oscillations in our data as well with $n^{1/4.3}$ as the oscillatory variable with $A=0.014$, $k=14.131$ and $\phi=-1.421$. For reference, the DG values are $A_{DG} =0.006$, $k_{DG}=11.416$ and $\phi_{DG} = -1.822$. However since \cref{eq:3d-2} doesn't conform with our data analysis strategy, we do not comment any further on this.
\end{remark}

\subsection{$d=4$ case}

There are $5$ basis functions in \cref{eq:asymptotic} in $d=4$. Fitting all $5$ basis functions to the data MC-4D-15-40-15000-delta.csv gives $\kappa = 2147$ just at $n_{min}=n_0+1 = 41$ while fitting the leading $4$ basis functions gives $\kappa = 336.2$ at $n_{min} = 41$. Thus, we fit only the leading $3$ basis functions. We take $\kappa_0 = 200$ which keeps $n_1 = 964$. The results are:
\be
    \alpha_1^{(4)} = 1.7215 \pm 0.0045 \\
    \alpha_2^{(4)} = 0.170 \pm 0.064 \\
    \alpha_3^{(4)} = 0.26 \pm 0.24
\ee
Now from MacMahon generating function $M_4$ one has up to 4 significant digits after decimal:
\be
    \mu_1^{(4)} = 1.6614
\ee
We have:
\be
    \alpha_1^{(4)} > \mu_1^{(4)}
\ee
and the deviation is $Z = 13.4\sigma$.

\subsection{$d=5$ case}

There are $6$ basis functions in \cref{eq:asymptotic} in $d=5$. However fitting $6, 5,4$ leading functions to the data MC-5D-15-30-15000-delta.csv gives values of $\kappa$ as $28843, 3807,$ and $546.6$ respectively at $n_{min}=n_0+1 = 31$. Trying to fit leading $3$ functions saturates $\kappa$ to $200$ prematurely at $n_1 = 511$. However if we try to fit leading $2$ functions, we need a very low value of $\kappa_0 = 17$ to stay within $n_1 \sim 900-1000$. A reduced chi squared goodness of fit test reveals that the $3$-parameter model is a better statistical fit than the $2$ parameter model at $n_{min}=31$. Hence following our strategy, we fit $3$ leading coefficients with $\kappa_0 = 200$ and $n_1 = 511$. The results are:
\be
    \alpha_1^{(5)} = 1.6521 \pm 0.0059 \\
    \alpha_2^{(5)} = 0.248 \pm 0.060 \\
    \alpha_3^{(5)} = 0.23 \pm 0.16
\ee
Now from MacMahon generating function $M_5$ one has up to 4 significant digits after decimal:
\be
    \mu_1^{(5)} = 1.5737
\ee
We have:
\be
    \alpha_1^{(5)} > \mu_1^{(5)}
\ee
and the deviation is $Z = 13.3\sigma$.

\subsection{$d=6$ case}

There are $7$ basis functions in \cref{eq:asymptotic} in $d=6$. However fitting $7,6, 5,4$ leading functions to the data MC-6D-15-21-15000-delta.csv gives values of $\kappa$ as $413152, 52350, 6410,$ and $808.7$ respectively at $n_{min}=n_0+1 = 22$. Trying to fit leading $3$ functions saturates $\kappa$ to $200$ prematurely at $n_1 = 250$. However if we try to fit leading $2$ functions, we need a very low value of $\kappa_0 = 20$ to stay within $n_1 \sim 900-1000$. A reduced chi squared goodness of fit test reveals that the $3$-parameter model is a better statistical fit than the $2$ parameter model at $n_{min}=22$. Hence following our strategy, we fit $3$ leading coefficients with $\kappa_0 = 200$ and $n_1 = 250$. The results are:
\be
    \alpha_1^{(6)} = 1.652 \pm 0.021 \\
    \alpha_2^{(6)} = -0.07 \pm 0.17 \\
    \alpha_3^{(6)} = 0.89 \pm 0.33 
\ee
Now from MacMahon generating function $M_6$ one has up to 3 significant digits after decimal:
\be
    \mu_1^{(6)} = 1.509
\ee
We have:
\be
    \alpha_1^{(6)} > \mu_1^{(6)}
\ee
and the deviation is $Z = 6.81\sigma$.

\section{Some concluding remarks}

Thus, the current state of comparison between leading order asymptotics of partitions with MacMahon numbers are as follows:
\begin{itemize}
    \item For $d=1,2$, they are identical, so same asymptotics. 

    \item For $3 \leq d \leq 6$, our work numerically shows that partitions grow strictly faster than MacMahon numbers and the deviations are significant in each dimension. Numerical evidence for $d=3$ partitions growing faster than MacMahon numbers was also provided by DG \cite{Destainville}. 

    \item For $d \geq 7$ it has been analytically shown that partitions grow strictly faster than MacMahon numbers \cite{Yeliussizov}. 
\end{itemize}

Our data analysis strategy is quite stringent and hence led to estimation of leading three asymptotic terms in all the dimensions and also quite conservative error-bars. Increasing the number of independent MCMC runs will reduce the $\delta_{\Delta}$ Monte Carlo statistical error leading to a much higher upper threshold for $\kappa_0$ which is currently at 200. Also, there might be other better data analysis strategies which can lead to narrower error-bars and help estimating other sub-leading terms. 

\acknowledgments

I would like to thank Suresh Govindarajan for introducing me to the problem. I would also like to thank Rudra Prosad Sarkar at Raman Research Institute for running my codes in his office desktop to generate the data.



\bibliographystyle{JHEP}
\bibliography{bibliography}      
\end{document}